\documentclass[nonacm,10pt,natbib=false]{acmart}

\renewcommand\footnotetextcopyrightpermission[1]{}
\usepackage[normalem]{ulem}

\usepackage[english]{babel}
\usepackage[utf8]{inputenc}

\usepackage{lscape}
\usepackage{stfloats}

\usepackage{pdfpages}

\usepackage{amsmath}
\usepackage{graphicx}
\usepackage[colorinlistoftodos]{todonotes}
\usepackage{hyperref}
\hypersetup{
	colorlinks   = true, % Removes boxes; colors links
	urlcolor     = black, % Color of external hyperlinks
	urlbordercolor=black, % Color of external links
	linkcolor    = blue, % Color of internal links
	linkbordercolor = white, % Hide underlines for internal links
	citecolor    = blue, % Color of citations
	citebordercolor = white, % Hide underlines for citations
	breaklinks   = true, % Allow links to break over lines 
}
\usepackage[hyphenbreaks]{breakurl}

\usepackage{breakurl}
\usepackage{array}
\usepackage[shortcuts]{extdash}
\usepackage{multicol,multirow}

\usepackage{fancyhdr}
\usepackage{pbox}

\usepackage[]{titlesec}
\titleformat*{\section}{\Large\bfseries\filcenter\scshape}
\titleformat*{\subsection}{\normalsize\bfseries}
\usepackage[skip=2pt,font=normalsize]{caption}

\usepackage{setspace}

\usepackage{geometry}

\usepackage{booktabs}
\newcolumntype{L}{>{\raggedright\arraybackslash}p} 
\newcolumntype{C}{>{\centering\arraybackslash}p} 
\newcolumntype{R}{>{\raggedleft\arraybackslash}p} 

\makeatletter
\let\@authorsaddresses\@empty
\def\blfootnote{\gdef\@thefnmark{}\@footnotetext}
\makeatother

\usepackage[
backend      = biber,
sorting      = none, % Sorts by order of appearance
citestyle    = numeric-comp, % Combine consecutive cites (e.g., [1-3])
maxbibnames  = 3, % Longer author lists are truncated to '1st author et al.'
mincrossrefs = 1000, % Lets crossrefs display correctly
]{biblatex} 

\bibliography{all.bib}

\DeclareSourcemap{
	\maps[datatype=bibtex, overwrite]{
		\map{
			\step[fieldset=editor, null]
			\step[fieldset=location, null]
			\step[fieldset=publisher, null]
			\step[fieldset=isbn, null]
			\step[fieldset=issn, null]
			\step[fieldset=pages, null]
		}
	}
}

\date{\today}
\title{Security policy audits: why and how}
\author{Arvind Narayanan, Kevin Lee} 
\affiliation{Princeton University\country{USA}}
\begin{document}

\maketitle             
%\vspace{1em}

\pagestyle{plain} 

{\bf Abstract.} Information security isn’t just about software and hardware — it’s at least as much about policies and processes. But the research community overwhelmingly focuses on the former over the latter, while gaping policy and process problems persist. In this experience paper, we describe a series of security policy audits that we conducted, exposing policy flaws affecting billions of users that can be — and often are — exploited by low-tech attackers who don’t need to use any tools or exploit software vulnerabilities. The solutions, in turn, need to be policy-based. We advocate for the study of policies and processes, point out its intellectual and practical challenges, lay out our theory of change, and present a research agenda.
\vspace{1em}

\begin{multicols}{2}

\noindent
{\bf \large Security policies matter, but you wouldn’t know it from conference proceedings\ 
\ 
}
\vspace{0.5em}

Most information security researchers would readily acknowledge that security isn’t just about software flaws: it’s also about policies and processes. Whether an organization uses multi-factor authentication, has an incident response plan, and trains employees to avoid phishing has as big an impact on security than whether the software it runs has exploitable bugs. Even the impact of bugs comes down to a policy question, namely the frequency of patching.

Yet, our research community is almost exclusively focused on technical questions. Studies relating to software and hardware vastly outnumber studies relating to organizational policies and processes. To quantify this, we reviewed all 114 accepted papers to the 2021 IEEE Symposium on Security and Privacy — one of the premier venues for information security research — and found 48 papers that were primarily about protocols and design, 54 about software, 6 about hardware, 5 about users, 1 about economics, and none that were primarily about policies.  

We think people don’t study policies and processes because it isn’t valued  in the community and because the methods aren’t well understood (a mutually reinforcing cycle). It’s much  easier to interact with software than opaque organizations, so that’s what researchers have focused on.

Over time, this gap has grown wide. Work on improving software and hardware security, while important, has a decreasing marginal impact. As Alex Stamos has pointed out, only a vanishingly small number of people are actually harmed by high-tech vulnerabilities \cite{stamos2019tackling}. Meanwhile, gaping policy and process problems persist. As a straightforward example, many, many companies have simply forgotten to put access controls on Amazon S3 buckets or other cloud storage, exposing the private data of millions~\cite{Leyden2022insecure}.

One approach to studying these questions is what we call a {\em security policy audit}. In this paper, we describe our experiences conducting a series of audits of user authentication policies of cellular service providers and websites. Our goals are to advocate for the study of policies and processes, point out its intellectual and practical challenges, lay out our theory of change, and present a research agenda for future security policy audits.

We use the term audit to refer to research done by researchers without companies' cooperation, unlike audits that companies regularly do, either internally or by contracting a security auditing firm. Why aren’t the regular kind of audits (and penetration tests) enough? Because the audits we describe aren’t done on behalf of a company but rather on behalf of the users or clients of the service it provides. Due to  misaligned incentives, companies often don't fix policy flaws that affect their users — sometimes billions of them, as we’ll see repeatedly. External audits aim to reveal these flaws, warn users of security risks, and  pressure  companies to change. Another difference is that the audits we describe are often exploratory and uncover new policy flaws, rather than being anchored in established criteria.

Security policy audits fall under the umbrella of human factors in security, and there is a strong connection to usable security. Unlike most usable security research, the objects of study aren’t end users, but companies, their policies, and processes. The human-computer interaction community has studied security practices {\em within} organizations \cite{mcgregor2017would,poller2017can}, so security policy audits might be a fruitful topic of collaboration between the security and HCI communities.\\

% \newpage

\noindent
{\bf \large Our work: exposing authentication flaws}
\vspace{0.5em}

We began this line of work in 2018 when we heard frequent reports of SIM swaps in the media. In a SIM swap, an attacker hijacks a victim’s phone number by posing as the victim and calling the carrier to request that service be transferred to a SIM card the attacker possesses. We wanted to look for systemic reasons behind the anecdotal evidence of SIM swaps being extremely common. We did so by attempting SIM swap attacks on ourselves: 10 attempts each at each of five prepaid wireless carriers in the United States~\cite{lee2020empirical}. 39 of our 50 attempts succeeded (including all 30 at the three major carriers AT\&T, Verizon, and T-Mobile) because carriers used insecure ways to authenticate the caller’s identity. For example, one carrier accepted details of recent incoming calls as evidence of identity, which can be trivially subverted by an attacker by calling the victim before attempting the SIM swap. In addition to bad policies, we stumbled upon bad processes, namely poor training of customer service representatives. Some reps simply forgot to authenticate us, others gave us hints or victims’ personal information, and yet others proceeded despite authentication failures!
 
During the course of this work, we happened to notice that attackers can achieve essentially the same effect by legitimately taking over someone’s number — if that person had relinquished their number. We learned that giving up one’s phone number is extremely common: 35 million numbers are disconnected every year in the United States. Carriers necessarily recycle these numbers to other subscribers because the pool of 10-digit numbers is finite. We had a hunch that people don’t always remember to unlink  numbers from online accounts before giving them up, which means that phone number takeover could lead to online account takeover.

We sampled 259 phone numbers available to new subscribers at two major carriers, Verizon and T-Mobile, and found that 171 of them were tied to existing accounts at popular websites~\cite{lee2021security}. Worse, about 40\% (100 of 259) of the numbers were linked to leaked login credentials on the web, which could enable account hijackings that defeat SMS-based multi-factor authentication. We also found that carriers failed to include any defensive measures in their recycling policies or online interfaces, such as rate limits for acquiring and relinquishing phone numbers, making the attacker’s task trivial.

We realized that the reason that SIM swap and number recycling attacks are so problematic is that SMS as a second factor for authentication is so prevalent, so we turned to that. We reverse-engineered the authentication policies of over 140 websites that offer phone-based authentication. Notably, we found 17 websites on which user accounts can be compromised based on a SIM swap alone, i.e., without a password compromise. Further, 83 websites defaulted to or recommended an SMS-based second factor over more secure options such as authenticator apps~\cite{lee2020empirical}.

Finally, to get a more complete picture of online account security, we turned to passwords. We examined the policies of 120 of the most popular websites for when a user creates a new password for their account. Despite well-established advice that has emerged from the research community, we found that only 13\% of websites followed all relevant best practices in their password policies. Specifically, 75\% of websites do not stop users from choosing the most common passwords --- like \texttt{abc123456} and \texttt{P@\$\$w0rd} --- while 45\%  burden users by requiring specific character types in their passwords for minimal security benefit. We found low adoption of password strength meters, a widely touted intervention to encourage stronger passwords, appearing on only 19\% of websites. Even among those sites, we found nearly half misusing them to steer users to include certain character classes, and not for their intended purpose of encouraging freely-constructed strong passwords~\cite{lee2022password}.

Table \ref{tab:overview} gives an overview of the studies we conducted.
\\ 

\noindent
{\bf \large Policy problems, policy solutions}
\vspace{0.5em}

All of these weaknesses were caused by flawed security policies and flawed enforcement of those policies. None were the result of software flaws. Attackers could remain within the functionality of the user interface and use the system with the same privileges as any other user, albeit with malicious intent. These low-tech attackers are {\em UI-bound adversaries}. Since the adversary doesn’t need to use any tools or exploit a software vulnerability, anyone could potentially be an attacker. Impersonating subscribers over the phone to request a SIM swap, intercepting SMS 2FA codes to hijack accounts, looking for vulnerable recycled numbers on number change interfaces, and guessing the password on online accounts are all examples of UI-bound adversaries. 

These unsophisticated attack vectors are considered less interesting by the research community, yet arguably cause the most harm. SMS-based authentication has been used for over two decades. Today, more secure user authentication methods like dedicated software apps and security keys exist, and their adoption has steadily increased. But SMS-based authentication is still widely used; according to a late-2021 survey, 85\% of people in the U.S. and U.K. have used SMS 2FA at least once in 2021, compared to 44.4\% for authenticator apps and 9.7\% for security keys~\cite{childers2021state}. Similarly, passwords remain the most common means of online authentication.

Call center authentication, such as for phone service, credit cards, health insurance, and utilities, is a low-tech authentication method involving no customer interaction with software. According to a 2021 industry report, however, call center fraud is the leading cause of account takeovers of financial accounts and the second-leading cause for non-financial accounts~\cite{neustar2021state}. Despite the insecurity, there has never been any academic research on call center authentication practices. Technological sophistication and research novelty are negatively correlated with security impacts on users.

Policy problems require policy solutions. We mean policy in the broadest sense. In the course of our research, we engaged with policymakers, carrier trade associations, and online services. We have also publicized our findings for users and journalists to learn about these weaknesses, to make informed decisions on their account security settings, and to educate others. 

We’d hoped that companies would fix their flawed policies when we presented our research to them. That happened in a few cases. One of the mobile carriers we studied informed us that it had partially implemented our recommendations on mitigating SIM swaps. Four of the 17 websites that were especially vulnerable to SIM swap attacks informed us that they had fixed the flaws in their authentication policy after our outreach. In light of our research on recycled numbers, the two carriers in our study have made changes to better inform their subscribers about re-assigning phone numbers. Our work also received a modicum of media attention. Journalists have an important role in security: their work can help users pick safer products and, in the long run, put pressure on companies to improve their security.

Yet, more often than not, we found that the market didn’t correct itself, or took too long. In fact, policy audits are so unusual that most companies have no idea how to deal with them, as opposed to their well-honed processes for receiving and acting upon software vulnerability disclosures~\cite{lee2020vulnerability}. This led to unexpected challenges with reporting security policy flaws. In our disclosure to the 17 websites in the SIM swaps study, five did not understand our vulnerability report despite our attempts to make it as clear as possible; three websites acknowledged SIM swap attacks but failed to realize that their authentication policy was allowing for vulnerable accounts, and the other two websites misrepresented our disclosure as feature requests. Moreover, four websites relied solely on third-party bug bounty platforms to be notified about vulnerabilities. Since members of the platform were only focused on reviewing software bugs, three of our four reports were dismissed. All in all, only four websites made changes in response to our disclosure. As for the wireless carriers, while we didn’t have trouble reporting our findings, only one of our five reports to the major carriers resulted in (partial) fixes.

If responsible disclosure and pressure from users do not lead to improvements to security practices, policymakers should require companies to make the improvements or to bear the consequences of noncompliance. We have had some success in influencing policy improvements that address the flaws we found. Most notably, in September 2021, the Federal Communications Commission (FCC) launched a formal rulemaking process to protect consumers from SIM swap and number portability attacks, citing our research on SIM swaps as justification~\cite{FCC-21-102}. Similar to how rulemaking in the 1960s started requiring safety measures in automobiles, government agencies can draw from research to make informed decisions to protect users in the digital space. \\

\noindent
{\bf \large Methods and challenges}
\vspace{0.5em}

So far we’ve talked about the importance and upside of doing policy audits. But it’s also important to be clear about the downsides. A major one is that there is no getting around the need for manual work. In some cases, we found that automation was impossible. In order to reverse-engineer authentication policies for SIM swaps, we had to be on the phone with customer service representatives (CSRs). In other cases, such as reverse engineering password policies, automation yielded poor results despite our best efforts. Crowd work wasn’t effective either, because understanding and executing the tasks required some security expertise.

In studying SIM swap attacks, we made a total of 50 calls (40 hours of work) to customer service at the five carriers. When checking how well websites stood up against SIM swap attacks, we created accounts at 145 websites, provided all requested personal information, and examined the 2FA / recovery option pairs that were available to us (90 hours of work). In studying recycled phone numbers, we logged available numbers on the number change interfaces at two major carriers (60 hours of work). We then checked whether 259 of the likely recycled phone numbers were still linked to existing profiles at six popular websites (10 hours of work). In studying password policies at 120 of the most popular websites, we reverse-engineered each website's blocklist strategy, composition rules, and password strength meter behavior. This required about 5,000 password change attempts (200-300 hours of work).

While repetitive, the work was not mindless. Tasks such as adapting to unexpected scenarios during customer service calls or reverse engineering a password policy require some creativity and ability to think on one’s feet. Besides, doing it manually leaves the researchers with a level of insight into the data that is hard to otherwise obtain, and leads to new research insights. In fact, at the beginning of data collection for our password policies study, we focused on only one of the three questions we eventually ended up studying; the other two made themselves apparent once we interacted with a few password change interfaces.

The challenge of reproducibility and consistency in manual data collection is an interesting one. In the SIM swap study, researchers who called the companies followed a script which was developed with various contingencies in mind. In the password policies study, a second researcher studied a sample of websites to ensure that the results were consistent.

Another challenge is that of generating quantitatively reliable measurements. Unlike software vulnerabilities, it is not enough to show that policy flaws exist. Fixing these flaws is rarely easy, and there are always trade-offs. So, for policy makers to act, it is important to quantify the scale of the problem in some way: e.g. the likelihood of the adversary succeeding or the number of users affected. Otherwise the affected companies could claim that they are one-off occurrences. In some cases we lucked out: since all 10 of our SIM swap attempts at each of the three major carriers were successful, we were able to demonstrate the seriousness of the problem despite a small sample size. In other cases we had to do careful statistical analyses, e.g. to estimate the number of available recycled phone numbers.

A final set of challenges is around ethics. Ethical questions are common in security research, but doubly so in studies involving interacting with humans and/or using real people’s data, as all of our studies did. We took many steps to minimize risk. Usually this was straightforward, but in some cases we had to come up with alternative research methods, and even abandoned some research questions because we couldn’t find a way to answer it using ethical methods. One limitation due to ethical considerations is that we studied only large companies. An audit of a 1-person startup might put undue burden on that individual, whereas in a large company with millions of users the burden is spread out among the employees, and is small relative to the benefit to the users.

In the United States, human-subjects research is overseen by Institutional Review Boards (IRBs). When interacting with IRBs, it’s important to remember that they have a narrow definition of human-subjects research and their understanding of this type of research is still evolving, which can lead to some frustration. For example, dozens of IRBs had already ruled out corporate policy and ``mystery shopper'' studies as human subjects research, yet our IRB said that customer service reps were human subjects in our SIM swap study. On the other hand, they ruled that if we obtained people’s recycled phone numbers and looked through their texts to see if they are receiving sensitive messages, that’s not human-subjects research! (We decided that this crossed an ethical line, so we did not do this; we only examined the metadata of communications using an automated service).\\ 

% \newpage
\noindent
{\bf \large A research agenda}
\vspace{0.5em}

In our work, we were able to explore only a tiny slice of the possibilities of security policy audits. In this section we’ll describe five major interrelated directions for future work that we think are promising. This isn’t comprehensive, but it should give you a good sense of the potential of this method. Note that there’s rarely an obvious way to conduct such an audit, so most of these research questions will require methodological innovation to execute while staying within ethical boundaries. But we’ll reference many studies below that have found such innovative methods, and we think it is possible more often than not.

The first research direction is authentication. Even within this space, our work only scratched the surface of what is possible. Password recovery procedures are often the Achilles’ heel of online authentication, and yet have never been audited at scale, to our knowledge. Besides, we did not study authentication policies at some of the most important websites, such as banks, because those require accounts with those institutions. Studying them would probably require crowd-sourcing. Similarly, we only studied call center authentication for wireless accounts. There is a vast set of other organizations that rely on call center authentication for sensitive business, including banks, credit card companies, government services, and hospitals.

Anecdotal evidence suggests that offline authentication is even more badly broken than online authentication. As security researchers, we can’t help notice it all around us, such as when a hotel clerk forgets to ask for ID when we request a replacement key~\cite{schneier2008giving}. Such failures can have severe consequences~\cite{waters2008room}. Yet systematic studies of offline authentication are practically nonexistent. Such studies would help convince companies and, when necessary, policy makers to act. 

That leads to our second research direction, which is on the integrity of physical systems. Voting infrastructure, the power grid, transportation systems, and many other physical systems cannot rely primarily on software to ensure integrity in the face of attacks, so policies and processes are paramount. Few of these systems have received any serious study in our community, notably election infrastructure, although again most of the research has focused on software and hardware. There are a couple of exceptions. A 2011 paper showed that “New Jersey’s protocols for the use of tamper-evident seals [on voting machines] have been not at all effective”~\cite{appel2011security}, and a 2014 paper found that “we find that [Estonia’s] I-voting system has serious architectural limitations and procedural gaps that potentially jeopardize the integrity of elections”~\cite{springall14security}. In most cases, however, security analysis of critical systems — say, the water supply~\cite{Clark2000protecting} — is conducted by other communities, when it happens at all. Information security researchers have a lot to bring to these research topics because of our security mindset.

That leads to our third research direction, which is on physical safety and security. It’s hard to think of a more important topic from a societal impact perspective. The set of technologies and policies that affect physical safety is vast; we’ll give a few examples. Survivors of intimate partner violence can regain some privacy and safety by leaving family phone plans, but do carriers’ policies make it possible for them to do so without the permission of the abuser? On gig economy platforms such as Uber, Lyft, and Airbnb, physical safety is a major concern. There have been many reports of inadequate (or, sometimes, overly harsh) preventive measures taken by the companies~\cite{Kerr2022more}, but few systematic studies. A final example: what happens to people’s online accounts when they die? If service providers don’t have policies to hand over control to the next of kin, it can lead to a loss of treasured memories, or finances, on top of the tragedy of death. In fact, in the U.S. service providers have a legal obligation to handle such requests appropriately, but there is very little information about what they actually do.

A notable success story of policy audits relating to physical safety comes from a recent study of the spyware used in intimate partner violence~\cite{Chatterjee2018spyware}. These are apps installed by abusers on victims’ phones, designed to hide their presence, enabling surveillance, harassment, control, and violence. The authors found that such tools are readily available through search engines and on app stores and that many of the tool developers promote their abuse. This study and related advocacy has led to companies tightening their policies and enforcement related to such spyware.

The fourth and related research direction is about privacy. Privacy auditing is an extremely active area of research, but most of it consists of technical audits, such as uncovering what data apps are sending to third parties. On the other hand, privacy policies are a rich source of information about apps’ privacy practices, but they are not directly useful to end-users. There have been a smattering of systematic studies of the content of privacy policies. Privacy policy audits neatly bridge these two types of studies: they enable uncovering a company’s actual, operative privacy policy, and see whether it matches its expressed policy. To some extent, aspects of privacy practices can be uncovered by technical analysis of apps~\cite{Slavin2016toward,zimmeck2019maps}, but in other cases it requires unusual methods such as posing as a potential commercial partner and requesting to buy data from the company of interest. Such methods have been used by journalists and privacy advocates~\cite{Cox2019gave} but not yet by researchers.

Privacy policy audits have uncovered some distressing findings. Facebook was caught soliciting phone numbers from users for two-factor authentication but then misusing them for targeted advertising~\cite{venkatadri2019investigating}, resulting in a \$5 bn settlement with the Federal Trade Commission. They have shone light on the dark world of data brokers. One study showed that over 90\% of U.S. Facebook users are successfully linked to data brokers, and that at least 40\% of data broker sourced user attributes are not at all accurate~\cite{Venkatadri2019auditing}.

The final research direction is about platforms —  search engines, social media, e-commerce,  gig economy marketplaces, etc.  Platforms have outsized power in our society and are inadequately regulated, especially in the United States. They have various security and safety related policies that inevitably have both false positives and false negatives. Often the policies themselves are not made clear; enforcement is almost always uneven. Audits are thus a vital accountability tool.

One set of platform policies that deserve scrutiny is around content moderation: removal or flagging of content that violates policies. A recent report by the U.S. Federal Trade Commission lists over a dozen categories: scammy content, deepfakes, fake reviews, sale of illegal substances, child sexual abuse, revenge pornography, cyberstalking, hate crimes, glorification of violence, incitement to violence, terrorists’ use of platforms, disinformation campaigns, and sale of counterfeit products~\cite{FTC2022combatting}. A few well-publicized concerns such as alleged political bias in disinformation removal have received some research attention, but the harms that everyday people suffer due to flawed policies and enforcement — such as weaponizing harassment reports or copyright complaints to try to get people banned — have largely flown under the radar. 

Another set of policies relates to advertising. We are gradually starting to get a picture of the problems with targeted advertising platforms, including discriminatory ads~\cite{Ali2019discrimination}, misleading or deceptive content, especially in the context of political advertising~\cite{Edelson2019analysis,Papakyriakopoulos2022algorithms}, and poor enforcement~\cite{Matias2022software,LePochat2022audit}, but there is a long way to go. In this area, the dearth of studies isn’t due to a lack of interest from researchers, but rather that platforms, notably Facebook, have been using legal and technical restrictions to actively inhibit researchers’ efforts to study them~\cite{NYUAdObservatory}.

Note that there is a burgeoning community of algorithm auditing that studies platform algorithms and policies~\cite{Costanza2022audits}. The algorithm auditing space is vast---it includes, for instance, revealing algorithmic bias~\cite{GenderShades} or pricing policies~\cite{Chen2015peeking}---but in the above paragraphs we've tried to identify the questions that are relevant to the security community. \\ 

\noindent
{\bf \large Final thoughts}
\vspace{0.5em}

We think that security policy auditing could be an entire subfield of research, rather than the occasional one-off studies we’ve seen so far. It can be at least as impactful as the body of work on improving software security. But keep in mind that it’s currently hard to publish security policy audits at top security venues. Reviewers at top conferences disproportionately merit work for publication based on factors like emerging technologies and sophisticated attack vectors. The kind of technology-enabled harm that happens in reality is the opposite: most attacks exploit current or older technologies, with attackers using the products as intended, albeit with malicious intent~\cite{stamos2019tackling,Mickens2014world}. We hope that this paper serves as a small first step towards changing this bias. We urge you to consider societal impact as a criterion in your own reviewing.
\vspace{1em}

{\bf Acknowledgment.} We conducted the research on which this paper is based in collaboration with Ben Kaiser, Jonathan Mayer, and Sten Sjöberg. We are grateful to Mihir Kshirsagar and Andrés Monroy-Hernández for useful feedback. The genesis of this paper was a talk given by Narayanan at the Cambridge security seminar.

\end{multicols}

% \includepdf{bigtable.pdf}
\begin{landscape}

\begin{table}[!htb]
    \centering
     \begin{tabular}{L{.065\textheight}L{.1\textheight}L{.16\textheight}L{.065\textheight}L{.046\textheight}L{.1\textheight}L{.15\textheight}L{.077\textheight}L{.083\textheight}L{.079\textheight}L{.13\textheight}}
    \toprule
    \footnotesize{{\bf Study}} & \footnotesize{\textbf{Research question}} & \footnotesize{\textbf{Finding}} & \footnotesize{\textbf{Gap}} & \footnotesize{\textbf{Hrs of work}} & \footnotesize{\textbf{Intellectual challenges}} & \footnotesize{\textbf{Ethical considerations}} & \footnotesize{\textbf{Attack type}} & \footnotesize{\textbf{Users affected}} & \footnotesize{\textbf{UI-bound adversary?}} & \footnotesize{\textbf{Root cause}} \\
    \midrule
    % Row 1
    \footnotesize{Lee et al. 2020} & \footnotesize{What are the authentication procedures that prepaid carriers use for SIM swaps?} & \footnotesize{All prepaid carriers studied used easily-obtainable information to authenticate callers.\vspace{1em} CSRs sometimes forgot to authenticate us, or guided our guesses, or proceeded despite us failing authentication.} & \footnotesize{Research ignores widely deployed authentication methods} & \footnotesize{40} & \footnotesize{Adapting to unexpected scenarios during calls.} & \footnotesize{We didn't record calls or note any identifying information about CSRs.\vspace{1em}\ \ \ \ \ \ \ \ \ \ \ \ \ \ \ \ \ \ \ \ \ \ \ \ \ \ \  We took over our own accounts that we had set up.} & \footnotesize{Actual} & \footnotesize{Millions} & \footnotesize{Yes} & \footnotesize{Hypothesis: carriers  largely externalize the costs of SIM swaps (online account takeover).} \\
    \midrule
    % Row 3

    % Row 4
    \footnotesize{Lee \& Narayanan 2021} & \footnotesize{What are the security and privacy risks of phone number recycling?} & \footnotesize{Most of the available phone numbers we sampled were recycled and also vulnerable to attacks against previous owners.\vspace{1em} \ \ \ \ \ \ \ \ \ \ \ \ \ \ \ \ \ \ \ \ \ \ \ \ \ \ \ \  Most number change interfaces had design weaknesses that could facilitate number recycling attacks.} & \footnotesize{Research ignores widely deployed authentication methods} & \footnotesize{70} & \footnotesize{Simulating a UI-bound adversary by creatively misusing features of various user interfaces.\vspace{1em} Statistical estimation techniques.} & \footnotesize{Avoiding harm to previous owners of recycled numbers---we aborted account recovery process after finding number was linked.} & \footnotesize{Hypothetical} & \footnotesize{Millions} & \footnotesize{Yes} & \footnotesize{10-digit phone numbers are finite and hence necessarily recycled.} \\
    \midrule
    % END Multirows
    % Row 5
    %% Col 1
    \footnotesize{Lee et al. 2020} &
    %% Col 2
    \footnotesize{How well do websites that offer SMS-based auth stand up against SIM swap attacks?} &
    %% Col 3
    \footnotesize{17 / 145 websites studied allowed SMS 2FA / recovery simultaneously.} &
    %% Col 4
    \footnotesize{Practice lags research} &
    %% Col 5
    %% Col 6
    \footnotesize{90} &
    %% Col 7
    \footnotesize{Exhaustively exploring each website's authentication logic.} &
    %% Col 8
    \footnotesize{We tested configurations on our own accounts that we had set up} &
    %% Col 9
    \footnotesize{Actual} &
    %% Col 10
    \footnotesize{Billions} &
    %% Col 11
    \footnotesize{Yes} &
    %% Col 12
    \footnotesize{Hypothesis: website operators view SIM swaps as someone else’s problem; they prioritize usability over security by overrelying on SMS 2FA.} \\
    \midrule
    % Row 6
    %% Col 1
    \footnotesize{Lee et al. 2022} &
    %% Col 2
    \footnotesize{Are top websites following best practices for encouraging stronger passwords?} &
    %% Col 3
    \footnotesize{Most websites are not following best practices from research.} &
    %% Col 4
    \footnotesize{Practice lags research} &
    %% Col 5
    %% Col 6
    \footnotesize{200-300} & 
    %% Col 7
    \footnotesize{Formulating different types of passwords to fully reverse-engineer password composition policy.} &
    %% Col 8
    \footnotesize{We tested passwords that had been leaked in previous data breaches. These passwords had already been publicly available for a while, which reduces risk to victims.} &
    %% Col 9
    \footnotesize{Actual} &
    %% Col 10
    \footnotesize{Billions} &
    %% Col 11
    \footnotesize{Yes (for online attacks); No (for offline attacks).} &
    %% Col 12
    \footnotesize{Researchers have not performed outreach on their recommendations; websites have various reasons for not following long-established best practices.} \\
    \bottomrule
    
    \end{tabular}
    \caption{Overview of the security policy audits we conducted.}
    \label{tab:overview}
\end{table}
\end{landscape}

\begin{multicols}{2}
% \bibliographystyle{plain}
% \bibliography{all}
\printbibliography

@conference{stamos2019tackling,
    author = {Alex Stamos},
    title = {Tackling the Trust and Safety Crisis},
    year = {2019},
    address = {Santa Clara, CA},
    publisher = {USENIX Association},
    month = aug,
    url = {https://www.usenix.org/conference/usenixsecurity19/presentation/stamos},
    booktitle = {28th USENIX Security Symposium (USENIX Security 19)},
}

@online{Leyden2022insecure,
  title   = {{Insecure Amazon S3 bucket exposed personal data on 500,000 Ghanaian graduates}},
  author  = {Leyden, John},
  date    = {2022-01-06},
  url     = {https://portswigger.net/daily-swig/insecure-amazon-s3-bucket-exposed-personal-data-on-500-000-ghanaian-graduates},
  urldate = {2022-07-19},
  note    = {{The Daily Swig}},
}

@inproceedings{lee2020empirical,
	title = {An Empirical Study of Wireless Carrier Authentication for {SIM} Swaps},
	author = {Lee, Kevin and Kaiser, Benjamin and Mayer, Jonathan and Narayanan, Arvind},
	booktitle = {Sixteenth Symposium on Usable Privacy and Security (SOUPS 2020)},
	year = {2020},
	isbn = {978-1-939133-16-8},
	pages = {61--79},
	url = {https://www.usenix.org/conference/soups2020/presentation/lee},
	publisher = {USENIX Association},
	month = aug,
}

@inproceedings{lee2021security,
	author={Lee, Kevin and Narayanan, Arvind},
	title={{Security and Privacy Risks of Number Recycling at Mobile Carriers in the United States}},
	booktitle={{Proceedings of the 2021 APWG Symposium on Electronic Crime Research (eCrime)}},
	year={2021},
	pages={1--17},
	publisher={{Institute of Electrical and Electronics Engineers (IEEE)}},
	isbn={978-1-6654-8029-1},
	issn={2159-1245},
	doi={10.1109/eCrime54498.2021.9738792},
}

@inproceedings{lee2022password,
	title = {Password policies of most top websites fail to follow best practices},
	author = {Lee, Kevin and Sj{\"o}berg, Sten and Narayanan, Arvind},
	booktitle = {Eighteenth Symposium on Usable Privacy and Security (SOUPS 2022)},
	year = {2022},
	url = {https://www.usenix.org/conference/soups2022/presentation/lee},
	publisher = {USENIX Association},
	month = aug,
}

@online{neustar2021state,
  author  = {Neustar},
  title   = {{2021 State of Call Center Authentication}},
  url     = {https://www.cdn.neustar/resources/whitepapers/risk/neustar-state-of-call-center-authentication-2021.pdf},
  urldate = {2022-05-13},
}

@techreport{childers2021state,
    title={{State of the Auth 2021: Experiences and Perceptions of Multi-Factor Authentication}},
    author={Dave Childers},
    institution = {{Duo Labs}},
    year={2021},
    url = {https://duo.com/assets/ebooks/state-of-the-auth-2021.pdf},
}

@online{lee2020vulnerability,
    author = {Lee, Kevin and Kaiser, Benjamin and Mayer, Jonathan and Narayanan, Arvind},
    title   = {Vulnerability reporting is dysfunctional},
    date    = {2020-03-25},
    url     = {https://freedom-to-tinker.com/2020/03/25/vulnerability-reporting-is-dysfunctional/},
    urldate = {2022-04-27},
    note    = {Freedom to Tinker},
}

@online{FCC-21-102,
  author  = {{Federal Communications Commission}},
  title   = {{In the Matter of Protecting Consumers from SIM Swap and Port- Out Fraud}},
  subtitle = {{Notice of Proposed Rulemaking}}, 
  url     = {https://docs.fcc.gov/public/attachments/FCC-21-102A1.pdf},
  urldate = {2022-05-04},
}

@online{schneier2008giving,
  title   = {{Giving Out Replacement Hotel Keys}},
  author  = {Schneier, Bruce},
  date    = {2008-11-13},
  url     = {https://www.schneier.com/blog/archives/2008/11/giving_out_repl.html},
  urldate = {2022-07-20},
  note    = {{Schneier on Security}},
}

@online{waters2008room,
  title   = {{`Room key given to rapist': hotel guest}},
  author  = {Waters, Georgia},
  date    = {2008-10-29},
  url     = {https://www.brisbanetimes.com.au/national/queensland/room-key-given-to-rapist-hotel-guest-20081030-geai51.html},
  urldate = {2022-07-20},
  note    = {{Brisbane Times}},
}

@article{appel2011security,
    author = {Appel, Andrew W.},
    title = {Security Seals on Voting Machines: A Case Study},
    year = {2011},
    issue_date = {September 2011},
    publisher = {Association for Computing Machinery},
    address = {New York, NY, USA},
    volume = {14},
    number = {2},
    issn = {1094-9224},
    doi = {10.1145/2019599.2019603},
    journal = {{ACM Transactions on Information and System Security}},
    month = {9},
    articleno = {18},
    numpages = {29}
}

@inproceedings{springall14security,
    author = {Springall, Drew and Finkenauer, Travis and Durumeric, Zakir and Kitcat, Jason and Hursti, Harri and MacAlpine, Margaret and Halderman, J. Alex},
    title = {Security Analysis of the Estonian Internet Voting System},
    year = {2014},
    isbn = {9781450329576},
    publisher = {Association for Computing Machinery},
    address = {New York, NY, USA},
    doi = {10.1145/2660267.2660315},
    booktitle = {Proceedings of the 2014 ACM SIGSAC Conference on Computer and Communications Security},
    pages = {703–715},
    numpages = {13},
    location = {Scottsdale, Arizona, USA},
}

@article{Clark2000protecting,
    author = {Clark, Robert M. and Deininger, Rolf A.},
    title = {Protecting the Nation's Critical Infrastructure: The Vulnerability of U.S. Water Supply Systems},
    journal = {Journal of Contingencies and Crisis Management},
    volume = {8},
    number = {2},
    pages = {73-80},
    doi = {https://doi.org/10.1111/1468-5973.00126},
    year = {2000}
}

@online{Kerr2022more,
  title   = {{More Than 50 U.S. Gig Workers Murdered on the Job in Five Years}},
  author  = {Kerr, Dara},
  date    = {2022-04-06},
  url     = {https://themarkup.org/working-for-an-algorithm/2022/04/06/more-than-50-u-s-gig-workers-murdered-on-the-job-in-five-years},
  urldate = {2022-07-20},
  note    = {{The Markup}},
}

@inproceedings{Chatterjee2018spyware,  
    author={Chatterjee, Rahul and Doerfler, Periwinkle and Orgad, Hadas and Havron, Sam and Palmer, Jackeline and Freed, Diana and Levy, Karen and Dell, Nicola and McCoy, Damon and Ristenpart, Thomas},  
    booktitle={2018 IEEE Symposium on Security and Privacy (SP)},   
    title={The Spyware Used in Intimate Partner Violence},   
    year={2018},  
    volume={},  
    number={},  
    pages={441-458},  
    doi={10.1109/SP.2018.00061}
}

@inproceedings{Slavin2016toward,
    author = {Slavin, Rocky and Wang, Xiaoyin and Hosseini, Mitra Bokaei and Hester, James and Krishnan, Ram and Bhatia, Jaspreet and Breaux, Travis D. and Niu, Jianwei},
    title = {Toward a Framework for Detecting Privacy Policy Violations in Android Application Code},
    year = {2016},
    isbn = {9781450339001},
    publisher = {Association for Computing Machinery},
    address = {New York, NY, USA},
    doi = {10.1145/2884781.2884855},
    booktitle = {{Proceedings of the 38th International Conference on Software Engineering}},
    pages = {25–36},
    numpages = {12},
    location = {Austin, Texas},
}

@article{zimmeck2019maps,
    title={{MAPS: Scaling Privacy Compliance Analysis to a Million Apps}},
    author={Zimmeck, Sebastian and Story, Peter and Smullen, Daniel and Ravichander, Abhilasha and Wang, Ziqi and Reidenberg, Joel R and Russell, N Cameron and Sadeh, Norman},
    journal={{Proceedings on Privacy Enhancing Technologies}},
    volume={2019},
    pages={66},
    year={2019},
    doi = {10.2478/popets-2019-0037}
}

@online{Cox2019gave,
    title   = {{I Gave a Bounty Hunter \$300. Then He Located Our Phone}},
    author  = {Cox, Joseph},
    date    = {2019-01-08},
    url     = {https://www.vice.com/en/article/nepxbz/i-gave-a-bounty-hunter-300-dollars-located-phone-microbilt-zumigo-tmobile},
    urldate = {2022-07-20},
    note    = {Motherboard},
}

@article{venkatadri2019investigating,
    title={{Investigating sources of PII used in Facebook's targeted advertising}},
    author={Venkatadri, Giridhari and Lucherini, Elena and Sapiezynski, Piotr and Mislove, Alan},    
    journal={{Proceedings on Privacy Enhancing Technologies}},
    volume={2019},
    pages={227--244},
    year={2019},
    doi = {10.2478/popets-2019-0013}
}

@inproceedings{Venkatadri2019auditing,
    author = {Venkatadri, Giridhari and Sapiezynski, Piotr and Redmiles, Elissa M. and Mislove, Alan and Goga, Oana and Mazurek, Michelle and Gummadi, Krishna P.},
    title = {Auditing Offline Data Brokers via Facebook's Advertising Platform},
    year = {2019},
    isbn = {9781450366748},
    publisher = {Association for Computing Machinery},
    address = {New York, NY, USA},
    doi = {10.1145/3308558.3313666},
    booktitle = {The World Wide Web Conference},
    pages = {1920–1930},
    numpages = {11},
    location = {San Francisco, CA, USA},
}

@techreport{FTC2022combatting,
  author  = {{Federal Trade Commission}},
  title   = {{Combatting Online Harms Through Innovation}},
  date = {2022-06-16},
  url     = {https://www.ftc.gov/reports/combatting-online-harms-through-innovation},
  urldate = {2022-07-20},
}

@article{Ali2019discrimination,
    author = {Ali, Muhammad and Sapiezynski, Piotr and Bogen, Miranda and Korolova, Aleksandra and Mislove, Alan and Rieke, Aaron},
    title = {Discrimination through Optimization: How Facebook's Ad Delivery Can Lead to Biased Outcomes},
    year = {2019},
    issue_date = {November 2019},
    publisher = {Association for Computing Machinery},
    address = {New York, NY, USA},
    volume = {3},
    doi = {10.1145/3359301},
    journal = {Proceedings of the ACM on Human-Computer Interaction},
    month = {11},
    articleno = {199},
    numpages = {30},
}

@misc{Edelson2019analysis,
    doi = {10.48550/ARXIV.1902.04385},
    author = {Edelson, Laura and Sakhuja, Shikhar and Dey, Ratan and McCoy, Damon},
    title = {{An Analysis of United States Online Political Advertising Transparency}},
    publisher = {arXiv},
    year = {2019},
}

@misc{Papakyriakopoulos2022algorithms,
    doi = {10.48550/arXiv.2206.04720},
    author = {Orestis Papakyriakopoulos and Christelle Tessono and Arvind Narayanan and Mihir Kshirsagar},
    title = {{How Algorithms Shape the Distribution of Political Advertising: Case Studies of Facebook, Google, and TikTok}},
    publisher = {arXiv},
    year = {2022},
}

@article{Matias2022software,
    author = {Matias, J. Nathan and Hounsel, Austin and Feamster, Nick},
    title = {Software-Supported Audits of Decision-Making Systems: Testing Google and Facebook's Political Advertising Policies},
    year = {2022},
    issue_date = {April 2022},
    publisher = {Association for Computing Machinery},
    address = {New York, NY, USA},
    volume = {6},
    doi = {10.1145/3512965},
    journal = {Proceedings of the ACM on Human-Computer Interaction},
    month = {4},
    articleno = {118},
    numpages = {19},
}

@inproceedings{LePochat2022audit,
    title = {{An Audit of Facebook's Political Ad Policy Enforcement}},
    author={Le Pochat, Victor and Edelson, Laura and Van Goethem, Tom and Joosen, Wouter and McCoy, Damon and Lauinger, Tobias},
    booktitle = {31st USENIX Security Symposium (USENIX Security 22)},
    year = {2022},
    address = {Boston, MA},
    url = {https://www.usenix.org/conference/usenixsecurity22/presentation/lepochat},
    publisher = {USENIX Association},
    month = aug,
}

@misc{NYUAdObservatory,
    title={{NYU Ad Observatory}},
    url={https://adobservatory.org/},
    urldate={2022-07-20},
}

@inproceedings{Costanza2022audits,
    author = {Costanza-Chock, Sasha and Raji, Inioluwa Deborah and Buolamwini, Joy},
    title = {Who Audits the Auditors? Recommendations from a Field Scan of the Algorithmic Auditing Ecosystem},
    year = {2022},
    isbn = {9781450393522},
    publisher = {Association for Computing Machinery},
    address = {New York, NY, USA},
    doi = {10.1145/3531146.3533213},
    booktitle = {{2022 ACM Conference on Fairness, Accountability, and Transparency}},
    pages = {1571–1583},
    numpages = {13},
    location = {Seoul, Republic of Korea},
}

@misc{GenderShades,
    title={{Gender Shades}},
    url={http://gendershades.org/},
    urldate={2022-07-20},
}

@inproceedings{Chen2015peeking,
    author = {Chen, Le and Mislove, Alan and Wilson, Christo},
    title = {{Peeking Beneath the Hood of Uber}},
    year = {2015},
    isbn = {9781450338486},
    publisher = {Association for Computing Machinery},
    address = {New York, NY, USA},
    doi = {10.1145/2815675.2815681},
    booktitle = {{Proceedings of the 2015 Internet Measurement Conference}},
    pages = {495–508},
    numpages = {14},
    location = {Tokyo, Japan},
}

@article{Mickens2014world,
  author  = {James Mickens},
  journal = {;login: logout},
  title   = {{This World of Ours}},
  year    = {2014},
  url = {https://www.usenix.org/system/files/1401_08-12_mickens.pdf},
}

@article{mcgregor2017would,
    author = {McGregor, Susan E. and Watkins, Elizabeth Anne and Caine, Kelly},
    title = {{Would You Slack That? The Impact of Security and Privacy on Cooperative Newsroom Work}},
    year = {2017},
    issue_date = {November 2017},
    publisher = {Association for Computing Machinery},
    address = {New York, NY, USA},
    volume = {1},
    number = {CSCW},
    doi = {10.1145/3134710},
    journal = {{Proceedings of the ACM on Human-Computer Interaction}},
    month = {12},
    numpages = {22},
}

@inproceedings{poller2017can,
    author = {Poller, Andreas and Kocksch, Laura and T\"{u}rpe, Sven and Epp, Felix Anand and Kinder-Kurlanda, Katharina},
    title = {Can Security Become a Routine? A Study of Organizational Change in an Agile Software Development Group},
    year = {2017},
    isbn = {9781450343350},
    publisher = {Association for Computing Machinery},
    address = {New York, NY, USA},
    doi = {10.1145/2998181.2998191},
    booktitle = {Proceedings of the 2017 ACM Conference on Computer Supported Cooperative Work and Social Computing},
    pages = {2489–2503},
    numpages = {15},
    location = {Portland, Oregon, USA},
}

\end{multicols}

\end{document}